\documentclass[aps,prl,twocolumn,floatfix]{revtex4-2}
\usepackage[colorlinks]{hyperref}
\usepackage[utf8]{inputenc}
\usepackage{graphicx}
\usepackage{multirow}
\usepackage{amssymb}
\usepackage{amsmath}
\usepackage{microtype}
\usepackage{braket}
\usepackage{xcolor}
\usepackage{slashed}

\begin{document}
	\title{Implications of W-boson mass anomaly for atomic parity violation}
	\author{H. B. Tran Tan}
	\author{A. Derevianko}
	\affiliation{Department of Physics, University of Nevada, Reno, Nevada 89557, USA}
\begin{abstract}
We consider the implication of the recent measurement of the W-boson mass $M_W$ [Science \textbf{376}, 170 (2022)] for atomic parity violation  experiments. We show that the change in $M_W$ shifts the Standard Model prediction for the ${}^{133}$Cs nuclear weak charge to $Q_W({}^{133}{\rm Cs})=-72.85(6)$, i.e. by $5.5\sigma$ from its current value. This brings existing experimental result for $Q_W({}^{133}{\rm Cs})$ into an essential agreement with the Standard Model. Using our revised value for $Q_W({}^{133}{\rm Cs})$, we readjust constraints on physics beyond the Standard Model.
\end{abstract}
	
\maketitle

\textit{\underline{Introduction}} - Atomic parity violation (APV) remains a major mean for testing the electroweak sector of the Standard Model (SM) at low energy. Currently, the best APV result provides a confirmation for the SM prediction of the ${}^{133}$Cs nuclear weak charge at the level of 0.35\% accuracy~\cite{Wood1997}. Future APV experiments with expected accuracy $0.1-0.2\%$~\cite{Gomez2005,DeMille2008,Antypas2013,choi2018gain,aubin2013atomic,portela2013towards,Altuntas2018,Aluntas2018_2} may help resolve the tension between high-energy $Z$ pole measurements of $\sin^2\theta_W$ (here $\theta_W$ is the weak mixing angle)~\cite{Abe_2000,aleph2011} when extrapolated to the $\sim100$ GeV scale. APV experiments are also uniquely sensitive to new physics effects to which high-energy probes are insensitive. These new phenomena include extra $Z'$ boson with masses at the TeV scale~\cite{Marciano1990,Marciano1992,Porsev2009}, novel four-fermion contact interactions~\cite{Eichten1983,Young2007}, very weakly coupled ``dark boson'' with mass in the MeV range~\cite{Davoudiasl_PRL_2012,Davoudiasl_PRD_2012,Andreas2012}, as well as the axion and axion-like particles~\cite{Derevianko2016}.

The use of APV to constrain physics beyond the SM relies on precise measurement of the APV amplitude $E_{\rm PV}$, accurate theoretical calculation of the atomic structure factor $k_{\rm PV}$ needed for extracting the the nuclear weak charge $Q_W$, and exact knowledge of the SM prediction against which the extracted value is compared. For ${}^{133}$Cs, the only existing measurement of $E_{\rm PV}$ comes from the Boulder group with an uncertainty of $0.35\%$~\cite{Wood1997}, although a new experiment is being planned with an aim of a $0.2\%$ accuracy~\cite{Antypas2013,choi2018gain}. 

Early atomic calculations of $k_{\rm PV}$ for ${}^{133}$Cs at the level of 0.4\% uncertainty~\cite{Dzuba1985,DZUBA1989,Blundell1990,Blundell1992} gave a value of $Q_W$ that is $2.5\sigma$ away from the SM prediction. Later developments resulted in the inclusion of sub-1\% contributions from Breit and QED corrections and culminated in the most detailed coupled-cluster single double and valence triple calculation (CCSDvT) with an uncertainty of $0.27\%$ and a value for $Q_W$ in perfect agreement with the SM~\cite{Porsev2010}. A more recent reevaluation yielded a $Q_W$ which is 1.5$\sigma$ away from the SM value whilst raising the uncertainty back to 0.5\%~\cite{Dzuba2012}. The latest calculation~\cite{SahDasSoi2021-CsPNC,roberts2021comment,Sahoo2022} gives a result agreeing with Ref.~\cite{Porsev2010} with an uncertainty of 0.3\%. A new parity-mixed coupled-cluster approach to calculating $k_{\rm PV}$ is under development~\cite{Tran2022}, with a goal of reducing the uncertainty to $0.2\%$. 

In this paper, we discuss the final ingredient, namely, the SM value of the ${}^{133}$Cs nuclear weak charge $Q_W$. More specifically, we consider the new measurement of the $W$ boson mass~\cite{MW2022} and its implication for the evaluation $Q^{\rm SM}_W$ and the interpretation of APV results. By using the new value of $Q^{\rm SM}_W$ implied by Ref.~\cite{MW2022} and existing APV results for ${}^{133}$Cs, we readjust limits on the weak isospin-conserving parameter of vacuum polarization effects on gauge boson propagators and limits on the mass of the $Z'$ boson. Implications of the new $W$ boson mass measurement~\cite{MW2022} for other physics beyond the SM scenarios were considered in Refs.~\cite{lu2022electroweak,fan2022w,de2022impact,asadi2022oblique,bhaskar2022,heckman2022,athron2022,athron2022w,liu2022,addazi2022}.

\underline{\textit{Theory}} - The electroweak SM is described in terms of the $SU(2)_L\times U(1)_Y$ gauge group with corresponding vector fields $W^i_\mu$ ($i=1,2,3$) and $B_\mu$ with gauge couplings $g$ and $g'$. Spontaneous breaking of the electroweak gauge symmetry is effected by introducing a complex scalar Higgs doublet $\phi$ with a Lagrangian
\begin{equation}
    \mathcal{L}_\phi=(D_\mu\phi)^\dagger(D^\mu\phi)+\mu^2\phi^\dagger\phi +\frac{\lambda^2}{2}(\phi^\dagger\phi)^2\,,\label{eq:Higg_potential}
\end{equation}
where the covariant derivative is defined as
\begin{equation}
    {{D}_{\mu }}\phi \equiv\left( {{\partial }_{\mu }}+\frac{ig}{2}{{\sigma }_{i}}W_{\mu }^{i}+\frac{i{g}'}{2}{{B}_{\mu }} \right)\phi\,.
\end{equation}
Here, $\sigma_i$ are the Pauli matrices. 

For $\mu^2<0$, the potential~\eqref{eq:Higg_potential} has a minimum at $v=\sqrt{2}|\mu|/\lambda$, around which point $\phi$ may be transformed into a single real scalar field $H$ with vanishing vacuum expectation value. After such a transformation, one finds that the Lagrangian~\eqref{eq:Higg_potential} contains the following terms
\begin{equation}
    \mathcal{L}_\phi\supset\frac12M_H^2H^2+M_W^2W^{\mu-}W^+_\mu+\frac12M_Z^2Z^\mu Z_\mu\,,\label{eq:symm_break>Lag}
\end{equation}
where
\begin{subequations}\label{eq:weak_bosons}
\begin{align}
    W_{\mu }^{\pm }&\equiv \frac{W_{\mu }^{1}\pm iW_{\mu }^{2}}{\sqrt{2}}\,,\\
{{Z}_{\mu }}&\equiv \frac{gW_{\mu }^{3}-{g}'{{B}_{\mu }}}{\sqrt{{{g}^{2}}+{{{{g}'}}^{2}}}}\,,
\end{align}
\end{subequations}
are the charged $W$ boson and neutral $Z$ boson fields, and
\begin{subequations}
\begin{align}
    M_H&=\lambda v\,,\\
    M_W&=gv/2\,,\\
    M_Z&=\sqrt{g^2+g'^2}v/2\,,
\end{align}
\end{subequations}
are the masses of the Higgs boson, $W$ boson, and $Z$ boson, respectively.

The Higgs field breaks the $SU(2)_L\times U(1)_Y$ symmetry down to an $SU(2)_{\rm weak}$ symmetry of weak interactions mediated by the $W^{\pm}$ and $Z$ bosons (see Eqs.~\eqref{eq:weak_bosons}) and a $U(1)_{\rm elec}$ symmetry with electromagnetic interactions mediated by the photon field ${A}_{\mu }\equiv \left(gW_{\mu }^{3}-{g}'{B}_{\mu }\right)/\sqrt{{{g}^{2}}+{{{{g}'}}^{2}}}$. With this, the Lagrangian for the fermion fields $\psi_i$ reads
\begin{align}\label{eq:Fermion_potential}
  {{\mathcal{L}}_{F}}&=\sum\limits_{i}{{{{\bar{\psi }}}_{i}}\left[ i\slashed{\partial }-{{m}_{i}}\left( 1+\frac{H}{v} \right) \right]{{\psi }_{i}}} \nonumber\\ 
  &-\frac{g}{\sqrt{2}}\left(J^{\mu\dagger}_WW^+_\mu+J^{\mu}_WW^-_\mu+J_A^\mu A_\mu +J^\mu_ZZ_\mu\right)\,,
\end{align}
where $m_i$ is the fermion mass and $\slashed{\partial}\equiv\gamma^\mu\partial_\mu$. Definitions for the weak charged current $J^\mu_W$, the weak neutral current $J^\mu_Z$, and the electromagnetic current $J^\mu_A$ may be found, e.g., in Ref.~\cite{ERLER2013119}. For small momentum transfer $Q^2\ll M^2_{W,Z}$, the interaction terms in Eq.~\eqref{eq:Fermion_potential} reduce to the effective charged and neutral current interactions
\begin{subequations}
\begin{align}
    \mathcal{L}_{\rm CC}&=-2J^{\mu\dagger}_WJ_{W\mu}/v^2\,,\\
    \mathcal{L}_{\rm NC}&=-\cos^2\theta_WJ^{\mu}_ZJ_{Z\mu}/v^2\,,\label{eq:NC_Lag}
\end{align}
\end{subequations}
where $G_F\equiv 1/(\sqrt{2}v^2)=g^2/(2\sqrt{2}M_W^2)$ is the Fermi constant and $\theta_W=\tan^{-1}(g'/g)$ is the Weinberg angle. The neutral current~\eqref{eq:NC_Lag} gives rise to the spin-independent APV considered in this paper. For a more comprehensive review of low-energy electroweak experiments, see, e.g., Ref.~\cite{Kumar2013}.

From Eqs.~\eqref{eq:Higg_potential} and~\eqref{eq:Fermion_potential}, we see that the electroweak SM depends on the set of parameters $\{g,g',\mu^2,\lambda^2,m_i\}$. The values of these parameters cannot be derived algebraically from within the SM and can only be determined experimentally. For this purpose, it may be more convenient to measure other sets of derived quantities, such as $\{g,\theta_W,M_H,v,m_i\}$ or $\{M_Z,\alpha,M_W,G_F,m_i\}$, where $\alpha=e^2/(\hbar c)$ is the fine-structure constant. Our approach should be contrasted with global fits with $M_W=80,433.5\pm9.4\,{\rm MeV/c^2}$ which do not show an appreciable change in the value of $\sin^2\hat{\theta}_W$, but rather slight variations in a wide array of electroweak parameters~\cite{lu2022electroweak,fan2022w,de2022impact,asadi2022oblique}. We believe that our approach makes the role of $M_W$ more explicit and eliminates a potential bias.

Among these derived parameters, the $M_Z$, $G_F$, and $\alpha$ have the lowest experimental errors. Namely, $M_Z=91.1876(21)$ GeV was determined from the $Z$ line-shape scan~\cite{Schael2006}, $G_F=1.1663787(6)\times10^{-5}\,{\rm GeV}^{-2}$ was derived from muon lifetime~\cite{MuLan2011}, and $\alpha= 1/137.035999084(21)$ was obtained by combining measurements of the $e^\pm$ anomalous magnetic moment~\cite{Tatsumi2019} with measurements of the Rydberg constant and atomic masses with interferometry of atomic recoil kinematics~\cite{Bouchendira2011,Parker2018}. As a result, we keep these fixed in our analysis below.

The quantities $\theta_W$, $M_W$, $M_H$, and $m_i$ are generally less well constrained (except for $m_{e,\mu,\tau}$). The Weinberg angle, or more precisely, $\sin^2\theta_W$, are measured in a variety of schemes, depending on the energy scale, including low-energy APV~\cite{BOUCHIAT1982,Wood1997,Guena2003,Guena2005,Tsigutkin2009,Macpherson1991,Meekhof1993,Phipp1996,Vetter1995,Edwards1995,antypas2019}, parity violating (PV) neutrino scattering~\cite{allaby1987precise,blondel1990electroweak,Zeller2002,Akimov2017}, as well as various types of asymmetries in scattering and decay processes at low energy~\cite{PRESCOTT1979,wang2014measurement,Wang2015,Hasty2000,BEISE2005,ARGENTO1983245,HEIL19891,Souder1990,jefferson2018} and high energy~\cite{Aaltonen2013,Schael2013,aaboud2018,Schael2006,Abe_2000,Abe_2001,Aaltonen2016,Abazov2018}. $M_W$ is obtained in $W$-pair production or single-$W$ production at energy $Q\sim M_Z$~\cite{Aaltonen2013,Schael2013,aaboud2018}. Combining $\sin^2\theta_W$ and $M_W$ allows one to constrain $M_H$ and the top quark mass $m_t$ via~\cite{RoPP2020}
\begin{equation}
    M_W^2\sin^2\theta_W=\frac{A^2}{1-\Delta r}\,,\label{eq:MWA}
\end{equation}
where $A\equiv\sqrt{\pi\alpha/(\sqrt{2}G_F)}$ and $\Delta r$ includes loop corrections to $M_W$, which depend on $m_t$ and $M_H$. Alternatively, one may used direct experimental value of $m_t$ and $M_H$ to constrain $M_W$ and $\sin^2\theta_W$.

We note that there exist in the literature several different definitions for $\sin^2\theta_W$. At the tree level, one has
\begin{equation}
\sin^2\theta_W=1-\frac{M_W^2}{M_Z^2}=\frac{g'^2}{g^2+g'^2}\,.\label{sin2_tree}
\end{equation}
One may promote the first of Eqs.~\eqref{sin2_tree} to a definition of the renormalized $\sin^2\theta_W$ to all orders in perturbation theory (the so-called on-shell scheme). In this case, the radiative correction $\Delta r$ has a quadratic dependence on $m_t$ and may receive large spurious contributions from higher orders $O(\alpha m^2_t/M_W^2)$. A more popular approach promotes the second of Eqs.~\eqref{sin2_tree} to an $\overline{\rm MS}$ (modified minimal subtraction) prescription with the quantity
\begin{equation}
    \sin^2\hat{\theta}_W(\mu)\equiv\frac{\hat{g}'^2(\mu)}{\hat{g}^2(\mu)+\hat{g}'^2(\mu)}\,,
\end{equation}
where $\hat{g}'$ and $\hat{g}$ are $\overline{\rm MS}$ quantities and $\mu$ is an energy scale conventionally chosen to be $M_Z$. With this definition, the identity~\eqref{eq:MWA} becomes
\begin{equation}
    M_W^2\sin^2\hat{\theta}_W=\frac{A^2}{1-\Delta \hat{r}_W}\,,\label{eq:MWA_hat}
\end{equation}
where the radiative correction $\Delta\hat{r}_W$ now has no quadratic dependence on $m_t$. The $\overline{\rm MS}$ and on-shell definitions are related via a multiplicative coefficient $c(m_t,M_H)=1.0351(3)$~\cite{RoPP2020}. For APV, the quantity $\hat{s}^2_0\equiv\sin^2\hat{\theta}_W(0)$ is relevant.

In what follows, we consider the recent measurement of $M_W$ by CDF II whose result shows a significant $7\sigma$ tension with the SM~\cite{MW2022}. We shall assume that all other quantities except $\sin^2\hat{\theta}_W$ remain the same. By the virtue of Eq.~\eqref{eq:MWA_hat}, the shift in $M_W$ then implies a corresponding change in the value of $\sin^2\hat{\theta}_W(M_Z)$, which would show up in low-energy APV experiments via $\hat{s}^2_0$.

\underline{\textit{Calculations}} - The nuclear weak charge $Q_W$ arises as a parameter of the effective Hamiltonian describing the electron-axial-vector–nucleon-vector interaction via the exchange of the $Z$ boson. It receives coherent contributions from both protons and neutrons and may be written as~\cite{RoPP2020}
\begin{equation}
    Q_W=Z\left(1-4\hat{s}^2_0\right)-N\,,\label{QW}
\end{equation}
where $Z$ is the atomic number, $N$ is the number of neutrons, and $\sin^2\theta_W$ is evaluated at the APV momentum transfer of $Q\approx2.4$ MeV. The nuclear weak charge $Q_W$ enters into the effective parity-violating interaction between an electron and the nucleus, which is given by
\begin{equation}\label{eq:H_W}
    H_W=-\frac{G_F}{2\sqrt{2}}\gamma_5Q_W\rho({\bf r})\,,
\end{equation}
where $\rho(r)$ is the nuclear charge density. The interaction~\eqref{eq:H_W} mixes atomic states with opposite parities and thus gives rise to electric dipole transitions between two states with the same nominal parity, e.g., $6S_{1/2}\rightarrow7S_{1/2}$ in ${}^{133}$Cs. A measurement of the amplitude of such a transition leads to the extraction of the value $\hat{s}^2_0$ which may then be compared with the SM prediction.

The value of $\hat{s}^2_0$ as predicted by the SM may be related to the $W$ boson mass $M_W$ via Eq.~\eqref{eq:MWA_hat} and the running of the weak angle~\cite{Czarnecki1996,CZARNECKI1998,CZARNECKI2000}
\begin{align}
    \sin^2\hat{\theta}_W(Q^2)&=\kappa(Q^2)\sin^2\hat{\theta}_W(M_Z)\nonumber\\
    &=\frac{\kappa(Q^2)A^2}{M_W^2\left(1-\Delta \hat{r}_W\right)}\,.\label{eq:s0_MW}
\end{align}
In this paper, we assume the standard value of $\kappa(0)\approx1.03$. Eq.~\eqref{eq:s0_MW} then shows that $\hat{s}^2_0$ is inversely proportional to $M_W^2$ (strictly speaking $\Delta \hat{r}_W$ also depends on $M_W$ via $\sin^2\hat{\theta}_W(M_Z)$; however, since $\Delta \hat{r}_W\approx0.06994\ll 1$, we can safely ignore this dependence).

Eqs.~\eqref{QW} and~\eqref{eq:s0_MW} show the dependence of the SM value for $Q_W$ on the physical mass $M_W$. Since the value of $M_W$ is such that $\hat{s}^2_0\approx1/4$, the dependence is relatively weak for heavy atoms where $N$ is large. Nevertheless, the extraordinary accuracy of APV experiments means that the measured weak charge could be sensitive to variations in experimental value of $M_W$. It is worth noting also that the suppression due to neutron is absent in the case of proton, whose weak charge $Q_W(p)=-0.0719(45)$~\cite{jefferson2018} has a good sensitivity to $\hat{s}^2_0$. Thereby, renewed efforts on atomic hydrogen APV experiment would be of great interest as an independent indirect probe of $M_W$  mass. 

Let us now consider the most recent result from the CDF II experiment at Tevatron~\cite{MW2022} which found $M_W=80,433.5\pm9.4\,{\rm MeV/c^2}$, equivalent to a $7\sigma$ deviation from the SM model value of $M_W=80,357 \pm 6\,{\rm MeV/c^2}$. Ref.~\cite{MW2022} also obtained a value of $M_Z=91,192\pm7.5\,{\rm MeV/c^2}$ which is consistent with the world average of $M_Z=91,187\pm2.1\,{\rm MeV/c^2}$. Plugging these values and the current SM value of $Q^{\rm SM}_W\left({}^{133}{\rm Cs}\right)=-73.23(1)$~\cite{RoPP2020} into Eqs.~\eqref{QW} and~\eqref{eq:s0_MW} while assuming that all other parameters are unchanged, we find that the CDF II result for $M_W$ implies
\begin{equation}
    Q_W^{\rm CDF II}\left({}^{133}{\rm Cs}\right) = -72.92(6)\,.\label{result}
\end{equation}
We note that in deriving the result~\eqref{result}, we have used Eq.~\eqref{QW} which neglects higher-order radiative corrections such as those from $WW$, $ZZ$, and $\gamma Z$ box diagrams (see, e.g., Ref.~\cite{ERLER2013119} for a comprehensive review). Since these extra contributions are at the level of a few percents, effects of varying $M_W$ on these terms are negligible compared to lowest-level change~\eqref{result}.

\begin{figure}[htb!]
    \centering
    \includegraphics[width=\columnwidth]{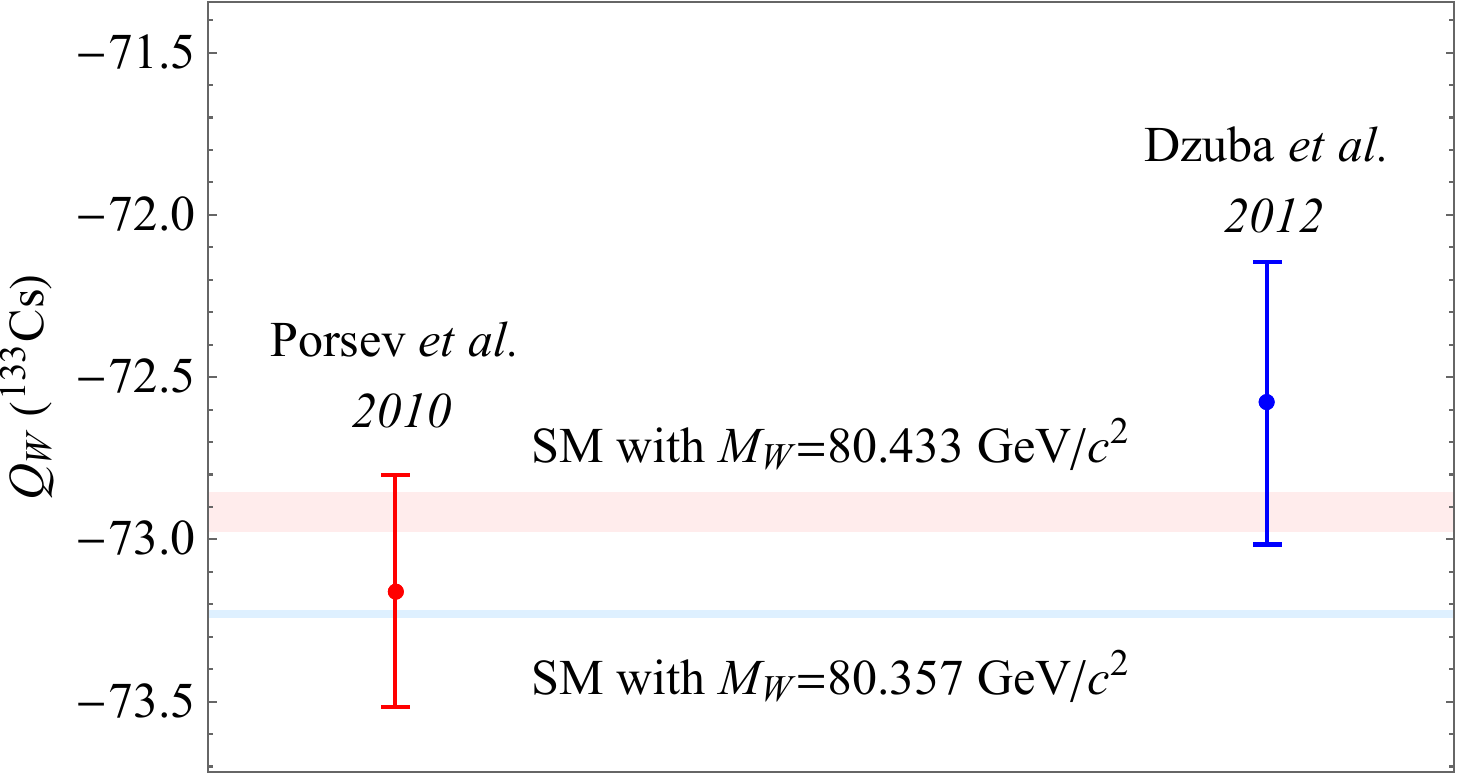}
    \caption{(Colors online) Comparison between the ${}^{133}$Cs nuclear weak charge as predicted by the Standard Model (SM) with mass of the $W$-boson being $M_W=80.357\,{\rm GeV/c^2}$~\cite{RoPP2020} (blue band), the (SM) with $M_W=80.433\,{\rm GeV/c^2}$~\cite{MW2022} (pink
    band), and ${}^{133}$Cs APV experiment~\cite{Wood1997} with different calculations for the ${}^{133}$Cs atomic structure factor (red and blue points).}\label{Lims}
\end{figure}

Our revised ${}^{133}{\rm Cs}$ weak charge~\eqref{result} corresponds to a shift of $-0.43\%$ or $5.4\sigma$ from the ``old'' SM value. Compared to the ${}^{133}{\rm Cs}$ weak charge extracted from Ref.~\cite{Porsev2010}, $Q_W^{\rm 2010} \left({}^{133}{\rm Cs}\right)=-73.16(29)_{\rm exp}(20)_{\rm th}$ and the result $Q_W^{\rm 2012}\left({}^{133}{\rm Cs}\right)=-72.58(29)_{\rm exp}(32)_{\rm th}$ extracted from Ref.~\cite{Dzuba2012} (the superscripts 2010 and 2012 denote the year of the corresponding publications), we have
\begin{subequations}
\begin{align}
    Q^{\rm 2010}_W-Q_W^{\rm CDF II} &= -0.25(36)\,,\label{eq:2010}\\
    Q^{\rm 2012}_W-Q_W^{\rm CDF II} &= 0.33(44)\,,\label{eq:2012}
\end{align}
\end{subequations}
where the deviations in Eqs.~\eqref{eq:2010} and~\eqref{eq:2012} were obtained by adding the corresponding theoretical and experimental errors and the uncertainty in Eq.~\eqref{result} in quadrature. From this, one observes that the result of Ref.~\cite{Porsev2010} is $0.7\sigma$ smaller than the CDF value while the result of Ref.~\cite{Dzuba2012} is $0.8\sigma$ larger. The comparison between these values are presented in Fig.~\ref{Lims}. Clearly, both result agree well with our revised SM value~\eqref{result} within their respective error bars, while their average is in excellent agreement with~\eqref{result}.

\textit{\underline{Discussions}} - We have demonstrated how the recent measurement of the $W$ boson mass~\cite{MW2022} meaningfully affects the interpretation of APV experiments. We find that the new value for the ${}^{133}$Cs nuclear weak charge reconciles the tension between the two most recent results for $Q_W\left({}^{133}{\rm Cs}\right)$ extracted from the same experiment~\cite{Wood1997} but with different methods of computing the atomic structure factor $k_{\rm PV}$.

The most substantial effect of the new value of $M_W$ is on APV in atomic hydrogen,
due to the absence of the otherwise leading and $M_W$-independent neutron contribution to nuclear weak charge. 
Indeed, $M_W=80,433.5\pm9.4\,{\rm MeV/c^2}$ implies a revised SM value for the proton weak charge
\begin{equation}
Q_W^{\rm CDF II}(p)\approx0.0800(16)\,,
\end{equation} 
corresponding to a shift of 12.5\% or $5.3\sigma$ away from the current SM prediction of $Q_W^{\rm SM}(p)=0.0711(2)$. It is also in a $\sim 2\sigma$ disagreement with the nuclear physics measurement $Q_W(p)=0.0719(45)$~\cite{jefferson2018}. A 1\% measurement of proton weak charge in APV experiments with hydrogen~\cite{Dunford2007,Rasor2020} may resolve this large difference.

An alternative APV approach are the measurements of APV in a chain of isotopes which forgoes evaluation of atomic structure factors $k_\mathrm{PV}$ altogether.  
Such measurements yield ratios of weak charges, $Q_W/Q_W^\prime$, of two nuclei with fixed number protons $Z$ but differing number of neutrons ($N$ and $N'=N +\Delta N$), see e.g. Ref.~\cite{antypas2019}. In the isotopic-chain measurements the sensitivity to $M_W$ (through $\hat{s}^2_0$) can be expressed as
$$
 \frac{Q_W/N}{ Q_W^\prime/N'} \approx 1  - \frac{\Delta N }{N} \frac{Z}{N} \left(1-4\hat{s}^2_0\right) \,
$$
while in a single-isotope measurement the relevant quantity is
$$
\frac{Q_W}{-N} = 1 - \frac{Z}{N} \left(1-4\hat{s}^2_0\right) \,.
$$
Comparing the two expressions we see that in the isotopic-chain measurements there is an extra suppression factor of $\Delta N/N$ which is $\ll 1$ for heavy nuclei of practical interest. Single-isotope measurements are more sensitive to  varying $M_W$ than the isotopic-chain experiments.


Finally, we note that the deviation between the values of the nuclear weak charges measured with Cs APV from their SM values have been used to constrain a variety of physics beyond the SM. In particular, Refs.~\cite{Porsev2010} and~\cite{Dzuba2012} considered new physics originating through vacuum polarization correction to gauge boson propagators as described by weak isospin conserving $S$ and isospin breaking $T$ parameters~\cite{Rosner2002}
\begin{equation}
    \Delta Q_W=Q_W-Q_W^{\rm SM} = -0.800S-0.007T\,,
\end{equation}
and placed limits on the parameter $S$. The new value for $Q_W^{\rm SM}$~\eqref{result} leads to a readjustment of these constraints, at the $1\sigma$ level, as
\begin{subequations}
	\begin{align}
		S^{2010}&=0.31(44)\,,\\
		S^{2012}&=-0.42(55)\,,
	\end{align}
\end{subequations}
where again the superscript 2010 corresponds to Ref.~\cite{Porsev2010} and the superscript 2012 corresponds to Ref.~\cite{Dzuba2012}. 

Additionally, a positive pull in $Q_W$ could  be indicative of an extra $Z'$ boson in the weak interaction~\cite{Marciano1990,Marciano1992,Porsev2009}
\begin{equation}
	\Delta Q_W\approx 0.4(2N+Z)(M_W^2/M_{Z'_\chi}^2)\,,
\end{equation}
where $M_{Z'_\chi}$ is the mass of the new $Z'$ boson. Using Eq.~\eqref{eq:2012} and $M_W=80,433.5\pm9.4\,{\rm MeV/c^2}$, we obtain \begin{equation}
    M_{Z'_\chi}\gtrsim1.3\,{\rm TeV}/c^2\,.
\end{equation}

\textit{\underline{Acknowledgements}} - The authors thank J. Erler and  I. Samsonov for helpful discussions. This work was supported in part by the U.S. National Science Foundation grant PHY-1912465, by the Sara Louise Hartman endowed professorship in Physics, and by the Center for Fundamental Physics at Northwestern University.

\bibliographystyle{apsrev4-2}
\bibliography{Lit}

\end{document}